# Theory of the force of Friction Acting on Water Chains Flowing through Carbon Nanotubes

J. B. Sokoloff[1,2] and A. W. C. Lau[2], 1. Northeastern University, Boston, MA 02115, 2. Florida Atlantic University, 777 Glades Rd, Boca Raton, FL 33431

A simple model for the friction experienced by the one dimensional water chains that flow through subnanometer diameter carbon nanotubes is studied. The model is based on a lowest order perturbation theory treatment of the friction experienced by the water chains due to the excitation of phonon and electron excitations in both the nanotube and the water chain, as a result of the motion of the chain. On the basis of this model, we are able to demonstrate how the observed flow velocities of water chains through carbon nanotubes of the order of several centimeters per second can be accounted for. If the hydrogen bonds between the water molecules are broken (as would occur if there were an electric field oscillating with a frequency equal to the resonant frequency of the hydrogen bonds present), it is shown that the friction experienced by the water flowing in the tube can be much smaller.

## I.    Introduction

Understanding the physics at the nanoscale for device applications, such as filtration devices [1-3], fuel cells [4-7] and sensing, devices [8], requires an understanding of water flow through a confined narrow tube-like structure. As is well known, classical hydrodynamics cannot account for the flow rate as measured in experiments and calculated in computer simulations of water flow in carbon nanotubes. In particular, the observed flow velocity of water through subnanometer diameter carbon nanotubes can be several orders of magnitude larger than that predicted by solutions of the Navier Stokes equation with perfect stick boundary conditions at the nanotube wall [9-11]. There have been many computer simulations addressing this problem [12-22], but a simple analytical treatment is still lacking. In this paper, we propose a simple analytical treatment of this problem based on the idea that water forms a one-dimensional water chain inside a confined narrow tube[17-21].

Recent molecular dynamics simulations show that water forms a one dimensional chain when it enters a sufficiently narrow carbon nanotube [12-21]. There is experimental evidence for this [22]. Simulations also show that water flows through aquaporins as one dimensional chains[23-27]. It is suggested that one reason for the drastic difference in the flow behavior of water in subnanometer diameter nanotubes may be the fact that it forms one dimensional chains [12-21].  An excellent illustration of the geometry of the water chain in a narrow channel can be found in Ref. 20.  Along the chain, each water molecule in the nanotube is hydrogen bonded to its two neighboring water molecules. In a hydrogen bond, the hydrogen atom on the bond is closer to one of the two oxygen atoms connected by the bond. In the ground state, each oxygen has two, and only two, hydrogen atoms near it. One lies on its hydrogen bond with one of its nearest neighbor molecules in the chain and one does not form a bond.

Presumably, the reason that water forms these one dimensional chains is that there is not enough room in the nanotube to accommodate bonding with additional water molecules that lie further from the symmetry axis of the nanotube. The total width of a 1-d chain is given by $b+2b\cos(54.5°)$, where $b$ is the radius of a water molecule (assuming that the water molecules are approximately spherical), since the angle between two hydrogen atoms belonging to a water molecule is about *109°*. If we assume that the water molecules in bulk water are nearly close-packed, $b$ is estimated to be *1.74A°*, which gives a diameter for the 1-d chain of 3.*76A°* compared with the tube diameter of *8.1A°*. (Incidentally, in the experiments of Ref. 9, the tube diameter is only *7nm*.) If we were to add an additional water molecule to the chain so it is hydrogen bonded to a dangling hydrogen bond on one of the chain's molecules, the added molecule will stick out an additional distance of *4bsin(54.5°)*. When this distance is added to the diameter of the chain, the total diameter becomes 10.56A°, which is larger than the *8.1A°* diameter of the narrow tubes.

There are two types of possible defects in a water chain: uncharged and charged defects. A pair of charged defects is produced by moving a hydrogen atom on one hydrogen bond from one side of the bond to the other, resulting in a molecule with three hydrogen atoms (a hydronium ion) and a neighboring molecule with only one hydrogen atom (a hydroxide ion). These are known as Bjerrum defects when they occur in ice, in which each water molecule is weakly hydrogen bonded to four neighboring water molecules. There can also be a defect in which a molecule in the chain is bonded to its neighboring molecules by hydrogen bonds, in which the hydrogens on the hydrogen bonds are either both close to the molecule or both close to one of its neighboring molecules. This type of defect has its dipole moment pointing perpendicular to the chain axis [20]. These defects have an energy of
$8 kcal/mole = 0.347 eV = 13.9 k_B T_{room}$, where $k_B$ is Boltzmann's constant and $T_{room}$ is room temperature. This implies that there are not likely to be that many of these defects excited at room temperature.

Therefore, the 1-d water chain is solid-like, over a long, but still finite, distance since 1-d solids do not have long-range order. Furthermore, the water chain should be located in the center of the nanotube because the rotational entropy is maximized if it resides there, as illustrated in Appendix A. The fact that the chain flows through the center of the tube minimizes the friction acting on the chain, resulting in the observed high flow velocity[9-11]. There have been several recent treatments of the friction experienced by water as it flows near a solid surface and through a confined geometry, such as Bocquet and Barrat [28], Huang and Szlufarska [29] and Kavokine, et. al. [30]. The treatments in these references are applicable to water in spaces of width greater than a nanometer, in which the water almost behaves as bulk water.

In this paper, we present a simple analytical theory to describe the flow of water through nanotube based on the concept of a water chain. Our treatment is specific to water in nanotubes of small enough diameters (i.e., subnanometer) so that the water flows through the tube as a chain of hydrogen bonded water molecules. We first derive, in Section II, the force of friction between the water chain and the carbon nanotube that results from phonon excitation of the water chain and the nanotube. The result suggests that the water chain experiences dry friction and a friction term that varies with flow velocity *v* as $1/|v|^{1/2}$, for large *v*. We then use these expressions for the friction to study the flow of the water chain, in Section III. Casting the problem of the motion of a water chain in terms of a Langevin equation, we derive the corresponding the Fokker-Planck equation, from which we calculate the (average) flow velocity as a function of an applied pressure. We find that using the parameters used in the molecular dynamics, the flow velocity is indeed enhanced in qualitative agreements with experiments. We also discuss the variation of the flow velocity as function of applied pressure. In Section IV, we consider the case in which a water chain was broken up into separate water molecules by an oscillating electric field. In this case, we calculate the flow of water molecules that are not hydrogen bonded together, and find that the flow velocity for a given applied pressure difference is increased over that for a water chain, in qualitative agreement with computer simulations.

## II.  Friction Experienced by a Water Chain Flowing through a Nanotube

There are two mechanisms for the friction that acts on a water chain flowing in a carbon nanotube, friction due to electronic excitations and friction due to phonon excitations. The friction due to electronic excitations was discussed in Ref. [31,32]. It is proportional to the flow velocity of the chain. In this section, we will discuss mainly the friction due to phonon excitations. Our calculation gives for the force of friction $f$ acting on the water chain

$$f = -\left(\Delta + \alpha |v|^{-1/2}\right) sign(v) \qquad (1)$$

where $v$ is the flow velocity. Using values of parameters from the Lennard-Jones interaction between the water chain and the nanotube used in molecular dynamics calculations performed on water in subnanometer carbon nanotubes we obtained $\Delta \approx 1.49 \times 10^{-16} N$ and $\alpha = 4.22 \times 10^{-12} N(m/s)^{1/2}$. This expression for the $f$ will be used in section III to determine $v$ as a function of the force on the water chain due to the pressure difference between the two ends of the nanotube.

The periodic part of the wall-water interaction can be approximated by the Steele potential[33], which is based on the Lennard-Jones interaction. The Steele treatment applied to a 1-d chain of molecules interacting with an atom in the tube wall gives a periodic part (corrugation) of the potential due to the chain of the form

$$V(x,z) = \sum_Q W_Q(z) e^{iQx} \qquad (2)$$

where $Q$ is a reciprocal lattice vector of the chain, $x$ is a distance along the chain and $z$ is the distance from a molecule in the chain to the wall. The quantity $W_Q(z)$, which is just the Fourier transform of the Lennard-Jones interaction, is given by

$$W_Q(z) = a^{-1} 4\varepsilon \int_{-\infty}^{\infty} dx' e^{-iQx'} \left[\frac{\sigma^{12}}{(x'^2+z^2)^6} - \frac{\sigma^6}{(x'^2+z^2)^3}\right]. \qquad (3)$$

Here, we have replaced the "zig-zag" structure of the water chain by a linear chain but we will use for the value of $z$ the actual distance of a water molecule from the wall of the nanotube. This should give the correct order of magnitude for the corrugation potential. By contour integration

$$I_n = \int_{-\infty}^{\infty} dx' e^{-Qx'} \frac{1}{(x'^2+z^2)^n} = \frac{2\pi}{(n-1)!} \frac{d^{n-1}}{dx'^{n-1}} \left[\frac{e^{iQx'}}{(x'+iz)^n}\right]_{x'=iz}. \qquad (4)$$

For $n=3$, we obtain

$$I_3 = \pi e^{-Qz} \left[ \frac{Q^2}{8z^3} + \frac{3Q}{8z^4} - \frac{3}{8z^5} \right]. \tag{5}$$

Then, the in the limit of large values of $Qz$,

$$W_Q(z) \approx \frac{\pi \sigma^6 Q^2 \varepsilon}{4az^3} e^{-Qz}, \tag{6a}$$

and hence

$$V(x,z) \approx \sum_Q \frac{\pi \sigma^6 Q^2 \varepsilon}{4az^3} e^{-Qz} \cos(Qx). \tag{6b}$$

Therefore, the periodic force along the chain $-\partial V(x,z)/\partial x$ is proportional to $sin(Qx)$. where $Q$ assumed to be the smallest value of this reciprocal lattice vector. In the absence of defects, the net friction will be negligibly small because the periodicities of the wall and the water chain are incommensurate. The Lennard-Jones (LJ) interaction between the nanotube and the 1-d water is very weak, because the energy scale parameter $\varepsilon$ in the interaction is only about $0.0028 eV$ [12-21]. For a metallic nanotube, there can also be an electrical image interaction between the dipole moment of a water molecule and the nanotube which is shown in Appendix B to be about $0.00125 eV$. From Eq. (6b), the time dependent force felt by the tube as the water moves through it is proportional to $\sin(x+Qvt)$. Substituting [12-21] $\varepsilon = 0.0028 eV$, $\sigma = 3.14 A°, Q = 3(A°)^{-1}, a = 2A°, z = 4A° - (2A°)\cos(54.5°) = 2.82 A°$, we get $\varepsilon' = W_Q(z) = 2.18 \times 10^{-24} J$.

Reference [34] uses a model for friction consisting of two crystals moving relative to each other with some disorder in the interaction between them. For the problem of a water chain moving in a nanotube, if the circumference of the nanotube is smaller than the phonon mean free path, the nanotube behaves as a one dimensional solid. If the circumference is greater than the phonon mean free path, the phonons will not have a way of knowing that the circumference is finite. Therefore, it behaves as a two dimensional solid. The force due to Eq. (6b) has the form

$$F_j = -\partial V / \partial x_j = -\lambda_j \sin Q(x_j + vt), \tag{7}$$

where $\lambda_j = (\pi \sigma^6 Q^2 \varepsilon_j / 4az^3)e^{-Qz}$, where $\varepsilon_j$ is the value of $\varepsilon$ on the $j^{th}$ site (the $j$ dependence accounts for disorder). The displacement along the tube $x_j$ of a carbon atom in the nanotube satisfies Newton's second law,

$$m \frac{d^2 x_j}{dt^2} = \sum_{j'} D_{j,j'} x_{j'} - \lambda_j \sin Q(x_j + vt), \tag{8}$$

where $D_{j,j'}$ is the dynamical matrix for the atomic displacements of the nanotube. Using the Green's function $G_{j,j'}(t-t')$, which satisfies

$$m\frac{d^2}{dt^2}G_{j,j'}(t-t') - \sum_{j''}D_{j,j''}G_{j'',j'}(t-t') = \delta_{j,j'}\delta(t-t'), \qquad (9)$$

Eq. (8) can be written as

$$x_j = -\sum_{j'}\int dt' G_{j,j'}(t-t')(\lambda_{j'}/m)\sin Q(j'a + x_{j'} + vt'), \qquad (10)$$

where the Green's function

$$G_{j,j'}(t-t') = N^{-1}\sum_{k,\alpha}\int d\omega \frac{e^{ik(j-j')a}e^{-i\omega(t-t')}}{-\omega^2 + \omega_\alpha^2(k) + i\gamma_0\omega}, \qquad (11)$$

where $\omega_\alpha(k)$ is the dispersion relation of the $\alpha^{th}$ phonon mode of wave vector $k$ and $\gamma_0$ is its inverse lifetime[35].

Neglecting $x_j$ in the argument of the sine function on the right hand side of Eq. (10) and averaging over $\lambda_j$, we obtain for the contribution to the average force of friction acting on the nanotube wall due to the excitation of phonons in the nanotube $F_{av1}$, which satisfies

$$\begin{aligned}F_{av1}v &= -T_m^{-1}\int_0^{T_m}dt'\sum_j\langle\lambda_j\sin Q(ja+vt)\dot{x}_j\rangle \\ &= -T_m^{-1}\int_0^{T_m}dt\sum_j\langle\lambda_j\sin Q(ja+vt)\sum_{j'}\int dt'\dot{G}_{j,j'}(t-t')(\lambda_{j'}/m)\sin Q(j'a+vt')\rangle\end{aligned} \qquad (12)$$

with $T_m$, the time over which the friction force is averaged and $<\cdots>$ signifies the average over $\lambda_j$. If $\lambda_j$ were independent of $j$, we would have a periodic structure, the water chain, moving relative to another periodic structure, the nanotube, which are incommensurate with each other. Therefore, there would be zero friction acting between them. Since the driving frequency due to the sliding water chain $Qv$ is much smaller than the Debye frequency of the nanotube wall, only acoustic modes are excited, and hence, we can set $\omega_\alpha(\vec{k}) = ck$ (for simplicity taking the phonon velocity $c$ of the three acoustic phonon modes to be equal), and hence, we have

$$F_{av1} = \frac{N_c\lambda_0^2}{mc}\frac{b}{2\pi}QI, \qquad (13)$$

where $\lambda_0^2 = <(\lambda_j - <\lambda_j>)^2>$, $N_c$ is the number of water molecules in the chain and

$$I = \int_{-\omega_D}^{\omega_D}cdk\frac{\gamma_0 Qv}{(c^2k^2 - Q^2v^2)^2 + \gamma_0^2Q^2v^2} = \frac{1}{2i}\int_{-\omega_D/c}^{\omega_D/c}cdk\left[\frac{1}{c^2k^2 - Q^2v^2 - i\gamma_0 Qv} - \frac{1}{c^2k^2 - Q^2v^2 + i\gamma_0 Qv}\right].$$
(14)

Since it is a good approximation to replace the Debye frequency $\omega_D$ by infinity, we can perform the integral over k by contour integration. Setting $\omega' = ck$, the above integral becomes

$$I = \frac{1}{2i} \int_{-\infty}^{\infty} d\omega' \left[ \frac{1}{(\omega'-r_1)(\omega'-r_2)} - \frac{1}{(\omega'-r_3)(\omega'-r_4)} \right]$$
$$= \frac{1}{4i} \int_{-\infty}^{\infty} d\omega' \left[ \frac{1}{r_1-r_2}\left( \frac{1}{\omega'-r_1} - \frac{1}{\omega'-r_2} \right) - \frac{1}{r_3-r_4}\left( \frac{1}{\omega'-r_3} - \frac{1}{\omega'-r_4} \right) \right] \quad (15)$$

where the poles of the integrand are located at $\omega' =$
$r_1 = re^{i\theta/2}$, $r_2 = re^{i(\theta/2+i\pi)} = -re^{i\theta/2}$, $r_3 = re^{-i\theta/2}$, $r_4 = re^{i(-\theta/2+i\pi)} = -re^{-i\theta/2}$, where
$r = (Q^4 v^4 + \gamma_0^2 Q^2 v^2)^{1/4}$, $\theta = \arctan(\gamma_0/Qv)$. Let us close the contour in the upper half plane. Then, since the poles $r_1, r_4$ lie in the upper half plane, we obtain

$$I = \frac{\pi}{(2)^{1/2}(Q^4 v^4 + \gamma_0^2 Q^2 v^2)^{1/4}} \left[ 1 + \frac{Q|v|}{(v^2 Q^2 + \gamma_0^2)^{1/2}} \right]^{1/2} \approx \frac{\pi}{(2)^{1/2} \gamma_0^{1/2} Q^{1/2} v^{1/2}} . \quad (16)$$

for small v. Therefore, from Eq. (13),

$$F_{av1} \approx \frac{N_c \lambda_0^2}{m} \frac{b}{2} Q \frac{1}{(2)^{1/2} c \gamma_0^{1/2} Q^{1/2} v^{1/2}} = \frac{N_c \varepsilon^2}{m} \frac{bQ^{1/2}}{2 \cdot 2^{1/2} a^2 c \gamma_0^{1/2} v^{1/2}} = \frac{\alpha}{v^{1/2}} , \quad (17)$$

where $\alpha = 4.22 \times 10^{-12} N(m/s)^{1/2}$ for the smallest water chain reciprocal lattice vector $Q \approx 2\pi/b$ and $\lambda_0 \approx \varepsilon'/b$, with $a = b = 2 \times 10^{-10} m$, for simplicity, with $\varepsilon' = 2.18 \times 10^{-24} J$, $m = 12(1.66 \times 10^{-27} kg) = 1.99 \times 10^{-26} kg$ and $N_c \approx 10^5$. The reason for using the value $N_c \approx 10^5$ is that the spacing between water molecules along the nanotubre is $2b \sin 54.5^o = 3.26 A^o$ for $b = 2A^o$. Then, for a 100 micron length nanotube (assuming that the water chains in Majumder, et. al.'s experiment[9] are of the order of the nanotube lengths, which are between 34 and 126 microns long, as shown in table S1 of the supplementary material for Ref. 9), the number of water molecules in the chain is $0.259 \times 10^6$. Our use of Lennard-Jones parameter $\varepsilon$, that was used in the simulations of Ref. 1 is somewhat arbitrary, as evidenced by the fact that the results of those simulations are so sensitive to its value[12]. Therefore, $\varepsilon$ should be thought of as a parameter that can be adjusted to give the observed results. Also, our model should be considered as a way to get a picture of the physical mechanisms responsible for the observed rapid flow of water through nanotubes and a way to predict possible qualitative trends in the problem, rather than device for getting accurate numerical results. For example, section IV gives a calculation of the mean velocity versus the applied force (due to the pressure difference across the tube) for unbonded water molecules. In that section, it is demonstrated that if the hydrogen bonds between water molecules in the chain are broken (e.g., by an oscillating electric field) the friction can be considerably smaller.

If the circumference of the tube is smaller than the phonon mean free path, the nanotube behaves as a two dimensional solid and

$$F_{j_1 j_2} = -\partial V / \partial x_{j_1 j_2} \approx -\lambda_{j_1 j_2} \sin Q(x^0_{j_1 j_2} + vt) \tag{18}$$

where $x_{j_1 j_2}$ is the x component of the displacement of the atom in the wall located at the point $j_1 \vec{a}_1 + j_2 \vec{a}_2$, where $\vec{a}_1, \vec{a}_2$ are the primitive lattice vectors of the atoms in the inner wall of the nanotube, where the x direction is the nanotube axis. The displacement $x_{j_1 j_2}$ satisfies the two dimensional version of Eq. (10)

$$x_{j_1 j_2} \approx -\sum_{j_1', j_2'} \int dt' G_{j_1 j_2, j_1' j_2'}(t-t')(\lambda_{j_1' j_2'}/m) \sin Q(x_{j_1' j_2'} + vt'), \tag{19}$$

where the Green's function

$$G_{j_1 j_2, j_1' j_2'}(t-t') = (N_1 N_2)^{-1} \sum_{\vec{k}, \alpha} \int d\omega \frac{e^{i\vec{k}\cdot(\vec{j}-\vec{j}')a} e^{-i\omega(t-t')}}{-\omega^2 + \omega_\alpha^2(\vec{k}) + i\gamma_0 \omega} \tag{20}$$

where $N_1$ and $N_2$ are the number of values of $j_1, j_2$, $\vec{j} = (j_1, j_2)$, $\omega_\alpha(\vec{k})$ is the dispersion relation of the $\alpha^{th}$ phonon mode of wave vector $\vec{k}$ and $\gamma$ is its inverse lifetime[24]. Replacing $x_{j_1 j_2}$ by $x^0_{j_1, j_2} = j_1 a_{1x} + j_2 a_{2x}$ and averaging over the values of $\lambda_{j_1 j_2}$, we obtain for the contribution to the average force of friction acting on the nanotube wall due to the excitation of phonons in the nanotube $F_{av2}$ from

$$F_{av2} v = -T_m^{-1} \int_0^{T_m} dt' \sum_{j_1 j_2} \left\langle \lambda_{j_1 j_2} \sin Q(x^0_{j_1 j_2} + vt) \dot{x}_{j_1 j_2} \right\rangle$$

$$= -T_m^{-1} \int_0^{T_m} dt \sum_{j_1 j_2} \left\langle \lambda_{j_1 j_2} \sin Q(x^0_{j_1 j_2} + vt) \sum_{j_1', j_2'} \int dt' \dot{G}_{j_1 j_2, j_1' j_2'}(t-t')(\lambda_{j_1' j_2'}/m) \sin Q(x^0_{j_1' j_2'} + vt') \right\rangle$$

(21)

Eq. (21) becomes on performing the average over $\lambda_{j_1 j_2}$

$$F_{av2} v = \frac{N_c \lambda_0^2}{m N_1 N_2} \sum_{\vec{k}, \alpha} \frac{\gamma_0 Q^2 v^2}{(Q^2 v^2 - \omega_\alpha^2(\vec{k}))^2 + \gamma_0^2 Q^2 v^2} T_m^{-1} \int_0^{T_m} dt \sin^2(Qvt + x^0_{j_1 j_2}), \tag{22}$$

where $\lambda_0 = \left\langle (\lambda_{j_1 j_2} - <\lambda_{j_1 j_2}>)^2 \right\rangle^{1/2}$. For large $T_m$,

$$F_{av2} = \frac{N_c \lambda_0^2}{2 m N_1 N_2} \sum_{\vec{k}, \alpha} \frac{\gamma_0 Q^2 v}{(Q^2 v^2 - \omega_\alpha^2(\vec{k}))^2 + \gamma_0^2 Q^2 v^2} . \tag{23}$$

Again, since the driving frequency due to the sliding water chain $Qv$ is much smaller than the Debye frequency of the nanotube wall, only acoustic modes are excited, and hence, we can set $\omega_\alpha(\vec{k}) = ck$ (for simplicity taking the phonon velocity $c$ of the three acoustic phonon modes to be equal), and hence, we have

$$F_{av2} = \frac{N_c \lambda_0^2}{2m} \frac{2\pi(2)}{(2\pi)^2} \Omega \int_0^{\omega_D/c} kdk \frac{\gamma_0 Q^2 v}{(Q^2v^2 - c^2k^2)^2 + \gamma_0^2 Q^2 v^2}, \quad (24)$$

where $\Omega$ is the unit cell area of the innermost nanotube and the factor (2) represents the number of acoustic modes. Performing the integral, we obtain

$$F_{av2} = \frac{N_c \lambda_0^2}{4\pi m} \frac{Q\Omega}{c^2} \left[ \arctan\left(\frac{\omega_D^2 - Q^2v^2}{\gamma_0 Qv}\right) + \arctan\left(\frac{Qv}{\gamma_0}\right) \right] \approx \frac{N_c \lambda_0^2 a}{8mc^2}. \quad (25)$$

since the Debye frequency $\omega_D \gg Qv, \gamma_0$ and since $\Omega \approx a^2$, where $a$ is of the order of the lattice constant of the nanotube. Using the parameters given under Eq. (17), we obtain

$$F_{av2} \approx \frac{N_c \varepsilon'^2 a}{8mc^2 b^2} \approx 1.49 \times 10^{-16} N, \quad (26)$$

where we assumed that $N_c \approx 10^5$. Note that unlike the usual solution of the Navier-Stokes equation with perfect stick boundary conditions, which gives a force of friction proportional to the flow velocity, we get a force of friction proportional to $(Qv)^{-1/2}$ if the phonon mean free path is larger than the circumference, and dry friction if the mean free path is smaller than the circumference. Although the above treatment of the friction experienced by water chains moving in nanotubes was based on classical mechanics, it is shown in appendix C that it can also be derived from a quantum mechanical treatment.

The pressure difference across the nanotubes in Majumder, et. al.'s experiment[9] was *1 atm.* Then, the total force applied to the chain is given by
$F = \pi R^2 \Delta P = \pi(0.4 \times 10^{-9} m)^2 (10^5 Pa) = 0.502 \times 10^{-13} N$, where $\Delta P$ is the pressure difference across the tube. Ref. 10 using different experimental methods gives smaller values of the flow velocity for a given applied pressure difference.

The lowest order perturbation theory treatment of the contribution to the friction due to phonons excited in the water chain will be valid when

$$b^2 \gg N^{-1} \sum_j T_m^{-1} \int_0^{T_m} dt < x_j^2(t) >, \quad (27)$$

where *N* is the number of molecules in the chain and where

$$x_j(t) \approx m^{-1} \sum_{j'} \lambda_{j'} \int_{-\infty}^{\infty} dt' G_{jj'}(t-t') \sin Q[vt' + j'a], \quad (28)$$

where *Q* is the smallest reciprocal lattice vector of the nanotube, or

$$x_j(t) = m^{-1}N^{-1}\sum_{j'}\lambda_{j'}\sum_{\vec{k},\alpha}e^{ikb(j-j')}\,\text{Im}\,\frac{e^{iQvt}}{\omega_\alpha^2(k)-Q^2v^2-i\gamma_0 Qv} \qquad (29)$$

where N is the number of atoms in the chain. Then,

$$N^{-1}\sum_j T_m^{-1}\int_0^{T_m} dt <x_j^2(t)> = (1/2)\lambda_0^2\frac{b}{2\pi m^2 c}I \approx \lambda_0^2\frac{b}{2\pi m^2}\frac{\pi}{(2)^{1/2}c\gamma_0^{3/2}Q^{3/2}v^{3/2}}, \qquad (30)$$

where $I$ is the integral given in Eq. (16). Eq. (30) shows that the inequality in Eq. (36) holds for

$$v^{3/2} >> \frac{1}{(2)^{1/2}}\frac{b}{2c}\frac{\lambda_0^2}{\gamma_0^{3/2}Q^{3/2}m^2b^2} = \frac{\varepsilon'^2}{2\cdot 2^{1/2}cba^2\gamma_0^{3/2}Q^{3/2}m^2}. \qquad (31)$$

For example, we saw above that $\lambda_0 \approx \varepsilon'/a$. Since there is only short range order in one dimension, the phonon wave functions are localized within a region whose length is known as the localization length $\ell$, which is of the order of the phonon mean free path [35]. Since $\gamma_0^{-1}$ is the lifetime of a phonon, the mean free path is equal to $c/\gamma_0 = \ell$, and hence, $\gamma_0 = c/\ell$. Then, let us use the following values for the parameters:
$\varepsilon' = 2.18\times 10^{-24}J, b = 2\times 10^{-10}m, c = 10^4 m/s, \gamma_0 = 10^{10}s^{-1}, \lambda_0 = \varepsilon'/a.$ The value chosen for $\gamma_0$ corresponds to a $\ell = 10^{-6}m$, a reasonable value for the localization length. Substituting in Eq. (31), we find that lowest order perturbation theory is valid for $v >> 0.561 m/s$. We will see in the next section, however, that there is good reason to believe that either potential $\varepsilon'$ used in these calculations is too large to account for the friction acting on the water chains reported in Ref. 1, and hence, the value used for $\lambda_0/<\lambda_j> = [<(\lambda_j - <\lambda_j>)^2>]^{1/2}/<\lambda_j>$ was too large. This would result in the value of $v$ given by Eq. (31) at which the perturbation theory is expected to break down will be much smaller.

The friction mechanisms discussed in the manuscript in principle could be applied to water in any subnanometer diameter tube. The perturbation theoretic treatment used in our manuscript, however, would not be applicable for nanotubes made from materials other than graphene because of the expected additional roughness in the walls of other tubes, which could result in much stronger interaction between the water chain and the nanotube wall.

### III. Motion of the Water Chain Under an Applied Pressure

In order to determine the mean velocity as a function of the applied force due to the pressure gradient, determined above Eq. (27) $F = \Delta P\pi r^2$, we will solve the following Langevin equation[36,37], which provides a simple way to include effects of thermal energy on the water-nanotube system:

$$M\frac{dv}{dt} = -\Delta sign(v) - sign(v)\alpha/|v|^{1/2} -(\gamma+\gamma_e)v + F + g\xi(t) \qquad (32)$$

where $\Delta = F_{av1}$, $\alpha$ is given under Eq. (1) and $M$ is the mass of the water chain. The term $-\gamma v$ represents viscous friction due to the interaction of the chain with thermal vibrations of the nanotube and the interaction of the nanotube with the thermal vibrations of the chain[38]. The term $-\gamma_e v$, represents the contribution to the friction coefficient of the nanotube wall due to intraband excitation of electrons [31,32 (Appendix A)]. For a 60*nm* diameter semiconductor nanotube the friction coefficient found in these references is $\lambda_e = 1.74 Ns/m^3$. This would give $\gamma_e = \lambda_e(2\pi RL) = 4.37 \times 10^{-16} Ns/m$ for a tube of radius *R=0.4nm* and length *L=100nm*. It is shown in Appendix D, however, that $\lambda_e \propto \exp(-g_1/k_B T)$ and since the nanotube band gap $g_1$ is inversely proportional to the tube radius, the number of conduction band electrons for a 0.8nm diameter nanotube will be several orders of magnitude smaller than that for a *60nm* diameter tube. Consequently, $\gamma_e$ will be many orders of magnitude smaller than the above value. If the nanotube is metallic, however, it is shown in Appendix D that $\gamma_e$ can be a many orders of magnitude larger than our estimate above of $\gamma_e$ for a 0.8nm diameter semiconducting nanotube. The nanotubes with a small enough inner radius to result in water forming chains are either single walled tubes or multiwalled tubes with negligible coupling between the tubes. Because of the small diameters of the nanotubes used in the experiments reported in Ref. 9, single particle electron excitations should adequately describe the friction due to electron excitations[30].

The Fokker-Planck equation corresponding to Eq. (32) is [36,37]

$$\frac{dP(v,t)}{dt} = \frac{\partial}{\partial v}\left[\frac{\Delta}{M}sign(v) + \alpha/|v|^{1/2} sign(v) - \frac{F}{M} + \frac{(\gamma+\gamma_e)v}{M} + \frac{g^2}{2M^2}\frac{\partial}{\partial v}\right]P(v,t), \quad (33)$$

where *P(v,t)* is the probability distribution of the velocity. The quantity $g^2 = 2k_B T\gamma$, so that for when $\alpha = \Delta = F = \gamma_e = 0$, $<v^2> = 2k_B T/M$, the thermal average value of *v* is given by Boltzmann statistics. The solution to Eq. (33) is

$$P(v,t) = Z^{-1}\exp\left[\frac{-\tau|v|\Delta - 2\tau\alpha|v|^{1/2} + Fv\tau - (1/2)(1+r)Mv^2}{k_B T}\right], \quad (34)$$

where $r = 1 + \gamma_e/\gamma$, $\tau = M/\gamma$ and $Z = \int_{-\infty}^{\infty} dv P(v,t)$. Introducing the following dimensionless quantities: $\bar{v} = v/v_T$, $\bar{\Delta} = \tau v_T \Delta/(k_B T)$, $\bar{F} = \tau v_T F/(k_B T)$, $\bar{\alpha} = 2\tau v_T^{1/2}\alpha/(k_B T)$, where $v_T = (2k_B T/M)^{1/2}$,

$$P(v,t) = Z^{-1}\exp\left[-|\bar{v}|\bar{\Delta} - \bar{\alpha}|\bar{v}|^{1/2} + \bar{F}\bar{v} - (1+r)\bar{v}^2\right], \quad (35)$$

from which we find

$$<\bar{v}> = 2Z^{-1}v_T\int_0^{\infty} d\bar{v}\bar{v}\sinh(\bar{F}\bar{v})\exp\left(-\bar{v}\bar{\Delta} - (1+r)\bar{v}^2 - \bar{\alpha}\bar{v}^{1/2}\right), \quad (36)$$

with

$$Z = 2v_T \int_0^\infty d\bar{v}\, \cosh(\bar{F}\bar{v}) \exp\left(-\bar{v}\bar{\Delta} - (1+r)\bar{v}^2 - \bar{\alpha}\bar{v}^{1/2}\right). \tag{37}$$

Let us first consider some results for insulating nanotubes, for which r=0.

When $\bar{F}$ is much larger than the friction resulting from the excitation of phonons, the $-\gamma v$ friction dominates, leading to

$$\bar{v} \approx (1/2)\bar{F} \tag{38}$$

or

$$v/v_T \approx \frac{\tau v_T F}{2k_B T} \tag{39}$$

which is equivalent to the expected result

$$v \approx \frac{\tau}{M} F = \frac{F}{\gamma}. \tag{40}$$

For the value of $\Delta = F_{av2}$ given under Eq. (1) and in Eq. (27) and the value of $\alpha$ given under Eqs. (1) and (17), $\bar{\Delta} = 2.24 \times 10^{-4}$, $\bar{\alpha} = 13.4$ for $\tau = 9.1 \times 10^{-9} s$, as an example. Then, Eqs. (36) and (37) give the results shown in Fig. 1, which shows typical behavior the flow velocity as a function of the applied force on the water chain.

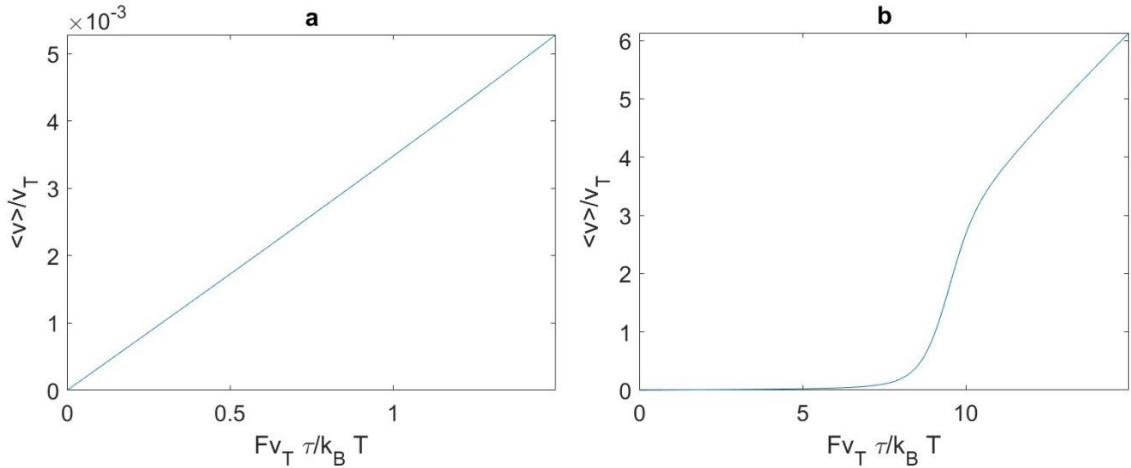

Figure 1: The dimensionless quantity $\bar{v}$ is plotted as a function of the dimensionless quantity $\bar{F}$ for $\bar{\Delta} = 2.24 \times 10^{-4}$, $\bar{\alpha} = 13.4$. a. for $0 < \bar{F} < 1.5$, b. $0 < \bar{F} < 15$.

We will use the plot of $\bar{v}$ vs. $\bar{F}$ to determine the value of F that results in a value of v comparable to the flow velocities found in Ref. 9. From Fig. 1b, we see that for $\bar{F} \approx 8$, v is approximately equal to $v_T$ (comparable to the experimental values reported in Ref. 9). This value of $\bar{F}$ corresponds to $F = 5.32 \times 10^{-12} N$, which is about two orders of magnitude larger than our estimate of $0.502 \times 10^{-13} N$ for F above Eq. (27). Although $\bar{\Delta}$ was included in these calculations for completeness, it makes a negligible contribution to the v vs. F curves.

The result of the same calculation for $\tau$ 2.5 times as large is given in Fig. 2. The value of $\tau$ used here is of the same order of magnitude as the value obtained from the viscous friction due to thermally activated fluctuations calculated by Volokitin[38], since he finds a friction coefficient $\lambda = 3 \times 10^3 Ns/m^3$, which gives $\gamma = 7.54 \times 10^{-13} Ns/m$, and hence, $\tau = M/\gamma = 2.64 \times 10^{-8} s$, using $M = 10^6 \times m = 1.99 \times 10^{-20} kg$.

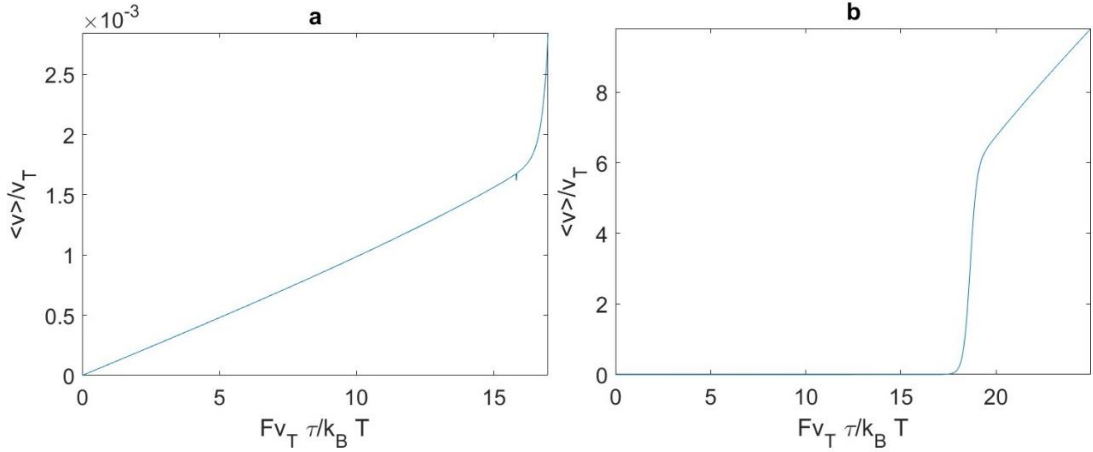

Figure 2: The dimensionless quantity $\bar{v}$ is plotted as a function of the dimensionless quantity $\bar{F}$ for $\bar{\Delta} = 5.6 \times 10^{-4}$, $\bar{\alpha} = 33.5$, a. for $0 < \bar{F} < 17$, b. for $0 < \bar{F} < 25$.

Since the damping constant in the viscous friction term $\gamma = M/\tau$, increasing $\tau$ results in a reduction of the viscous friction. We can see from Fig. 2b that $<v> = v_T$ when $\bar{F} \approx 18$, which corresponds to $F = 4.78 \times 10^{-12} N$, which is also 2 orders of magnitude larger than estimate above for F from the applied pressure difference in the measurements reported in Ref. 9.

If the water in the 0.8 nm nanotube could be described by the classical hydrodynamic treatment, in which the water is treated as a continuum with perfect stick boundary conditions at the tube wall, the flow velocity as a function of the applied pressure difference between the ends of the tube is given by

$$v = \frac{R^2 \Delta P}{8 \eta L} = \frac{F}{8 \pi \eta L}, \qquad (41a)$$

where $\eta$ is the viscosity of the water in the tube, R is the tube radius and L is the tube length. For $L \approx 100 \mu m$, the value that we have assumed for the nanotubes in Ref. 9,

$$\frac{v}{F} = 3.98 \times 10^5 m/Ns, \qquad (41b)$$

compared to $v/F = 1.33 \times 10^{11} m/Ns$ for the calculation described in the previous paragraph. The ratio of these two numbers $3.34 \times 10^5$ is the enhancement factor for the flow velocity for a 0.8nm diameter carbon nanotube for a given applied pressure, relative to the flow velocity found using classical continuum hydrodynamics.

If we double this value of $\tau$, we get the result shown in Fig. 3.

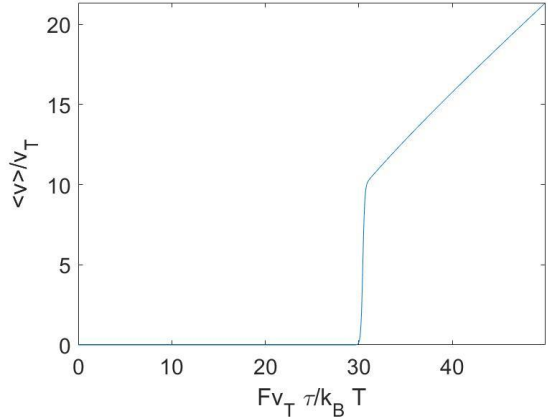

Figure 3: The dimensionless quantity $\bar{v}$ is plotted as a function of the dimensionless quantity $\bar{F}$ for the range $0 < \bar{F} < 45$, for $\bar{\Delta} = 5.6 \times 10^{-4}$, $\bar{\alpha} = 67$.

In this case, $v = v_T$ for $\bar{F} = 30$, which corresponds to $F = 3.98 \times 10^{-12} N$. For large enough values of $\bar{F}$, a typical plot of $\bar{v}$ vs. $\bar{F}$ has the shape of the curve shown in Fig 1b. As $\gamma$ becomes smaller, and hence $\tau$ becomes larger, this curve looks more like Fig. 2b or Fig. 3, i.e., more like a step function. The reason for this behavior is that as $\gamma$ is reduced, there is less thermal noise. Since both contributions to the phonon friction in Eq. (32) are opposite the direction of F, the water chain will not move in the direction of F unless there is a force due to thermal fluctuations in Eq. (32) in the direction of F, which exceeds $\alpha/|v|^{1/2} + \Delta - F$. As $\gamma$ is reduced, the probability of having a force that exceeds $\alpha/|v|^{1/2} + \Delta - F$ decreases, resulting in relatively small values of $<\bar{v}>$.

The fact that these calculations show that F is a couple of orders of magnitude larger than our estimate of the force due to the applied pressure above Eq. (22) implies that the value used for $\lambda_0/<\lambda_j> = [<(\lambda_j - <\lambda_j>)^2>]^{1/2}/<\lambda_j>$ was too large, either because $\varepsilon'$ was too large or there were fewer defects in the water chain and the nanotube. F vs. $\tau$ is plotted in Fig. 4.

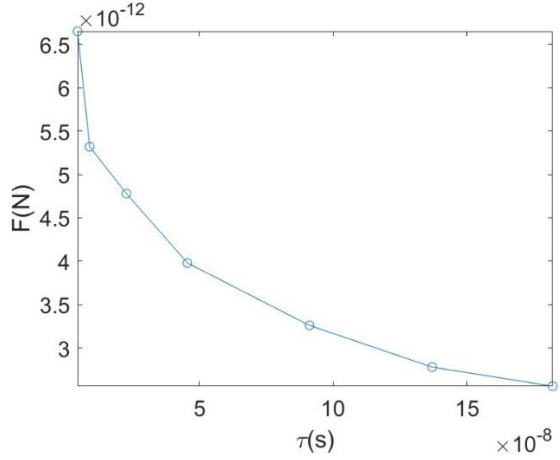

Figure 4: F is plotted as a function of $\tau$.

Although it appears that we can get results that are independent of $\gamma$ by taking the large $\tau$ (i.e., small $\gamma$) limit, since the plot of F vs. $\tau$ in Fig. 4 levels off as $\tau$ increases, we find that the v vs. F curve behaves more and more like a step function for larger values of $\tau$. This implies that v remains quite small until F reaches a critical value where the v vs. F curve increases sharply. This may suggest that the limit of vanishing viscous friction does not give a correct description of the problem. Rather, the viscous friction plays an important role in describing the flow of water in sub nanometer diameter carbon nanotubes. As suggested above, it could represent the friction due to roughness resulting from thermal fluctuations[38]. Since molecular dynamics simulations of water in carbon nanotubes use thermostats with relaxation times in the femtosecond range, compared to a relaxation time $\tau$ of the order of $10^{-8} s$ used in our calculations, the value of $\gamma$ due to the thermostat's relaxation time in the simulations will likely dominate over that due to the thermal noise that we assumed in our calculations. Consequently, one expects that the dependence of v on F will look more like Eq. (40) than the dependence due to friction resulting from phonon excitations that we reported here.

In the experiments reported in Ref. 10, $L = 8.02 \times 10^{-2} m$, and hence, $N_c = 2.47 \times 10^8$, $M = mL/b = 4.92 \times 10^{-18} kg$, $v_T (2k_B T/M)^{1/2} = 0.0404 m/s$, $\tau = M/\gamma = 0.653 \times 10^{-5} s$, using the value of $\gamma$ given above Fig. 2, which results in $\bar{\alpha} = 2.77 \times 10^3$ and $\bar{\Delta} = 0.983 \times 10^{-2}$. Since $v = 20 \mu m/s$, $\bar{v} = v/v_T = 4.95 \times 10^{-4}$. Since it is not possible to do the integrals in Eq. (36) and (37) for such large values of $\bar{\alpha}$, in order to estimate the value of F that gives $v = 20 \mu m/s$, we have calculated $\bar{v}$ versus $\bar{F}$ for several values of $\tau$ and then used the resulting plot of F versus $\tau$, shown in Fig. 5 to estimate the value of F for $\tau = 0.653 \times 10^{-5}$. Following this procedure, we estimate that $F \approx 10^{-11} N$. The value of $\Delta P$ estimated from Eq. (5) in Ref. 10, using the slip-length of 53nm quoted in table 2 in Ref. 10, is $1.51 \times 10^7 Pa$, giving $F = \Delta P \pi R^2 = 7.59 \times 10^{-10} N$. Then, our estimate for F is about one order of magnitude larger than the experimental value given in Ref. 10. Again, this could be due to the fact that the value of $\lambda_0$ that we use is too large, as discussed above after Fig. 2.

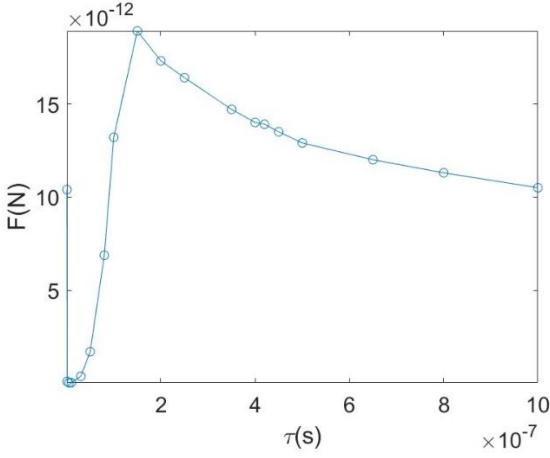

Fig. 5 : F is plotted as a function of $\tau$ corresponding to values the experimental of $\bar{v}$ and $\tau$ obtained from the experimental results reported in Ref.

From Fig. 5 a, we obtain $F \approx 10^{-11}$, which gives

$$\frac{v}{F} \approx 2 \times 10^6 \, m/Ns \qquad (42)$$

compared to $v/F = (8\pi\eta L)^{-1} = 4.96 \times 10^2 \, m/Ns$ for classical hydrodynamics with perfect stick boundary conditions, giving an enhancement factor of $0.4 \times 10^4$, which is comparable to the enhancement factors obtained in Ref. 10. The values of F to the right of the maximum in Fig. 5 correspond to solutions for F that occur in the part of the $\bar{v}$ versus $\bar{F}$ curve in which there is a steep rise, as shown in Figs. 1b, 2b and 3b, while the values of F to the left of the maximum correspond to the part of the $\bar{v}$ versus $\bar{F}$ curve to the left of the steep rise.

The singularity at v=0 in the friction force proportional to $|v|^{-1/2}$ should go away if we were to carry out the perturbation treatment to higher order in $\lambda_0$. Since the singularity is integrable, however, we can still from calculate <v> from this form for the friction. A reasonable interpretation of the values of <v> that we obtained from the solution of the Fokker-Planck equation is that since we have a distribution of chain velocities, there is a distribution of nonzero initial chain velocities.

## IV. Friction Acting on Single Water Molecules

There have been simulations that show that the flow rate of water through a carbon nanotube increases when the hydrogen bonds between water molecules get broken by an applied electric field oscillating with a frequency in resonance with and thus able to break up the hydrogen bonds[39,40]. In these simulations, the peak rate of flow of water molecules through the tube is $\phi = 37 ns^{-1} = 3.7 \times 10^{10} \, molecules/s$, and hence, the flow velocity is $b\phi = (2 \times 10^{-10} m)(3.7 \times 10^{10} s^{-1}) = 7.4 m/s$ when the hydrogen bonds are broken by the oscillating field. When the hydrogen bonds are intact, the flow rate is $3 ns^{-1}$, which gives a flow velocity of 0.6m/s. This reduction in the friction may be partly due to the fact that there are fewer water molecules in the tube at the same time when the water molecules are not bonded together, but it could also partially result from there being lower friction between individual water molecules and the nanotube. The latter possibility will be explored in this section. In order to determine the total force of friction acting on a collection of water molecules that are not bonded together, let us now consider the friction acting on an individual water molecule. We use the following procedure: For the case in which the nanotube behaves as a one dimensional solid, the displacement $x_j$ satisfies

$$x_j \approx -\sum_{j'} \int dt' G_{j,j'}(t-t')(\tilde{\lambda}/m)\delta(j'a - vt'), \qquad (43)$$

where $m$ is the mass of a carbon atom,

$$\tilde{\lambda}\delta(ja - vt) \qquad (44)$$

represents the force on the nanotube wall due to a single water molecule moving with a velocity $v$ and $G_{j,j'}(t-t')$ is the Green's function given in Eq. (11). Then, doing the integral over t' and the sum over j' in Eq (43) and using the fact that $\sum_{j'} e^{i(\omega/v - k)j'a} = N\delta_{k,\omega/v}$, we obtain

$$x_j = (\tilde{\lambda}/mv)\sum_{\alpha} \int_{-\infty}^{\infty} \frac{e^{ikja}e^{-i\omega t}}{(c^2/v^2 - 1)\omega^2 + i\gamma_0\omega} d\omega, \qquad (45)$$

and hence,

$$\dot{x}_j = 2(\tilde{\lambda}/mv)\sum_{\alpha} \int_{-\infty}^{\infty} d\omega \frac{i\omega e^{-i\omega t}e^{ikja}}{(c^2/v^2 - 1)\omega^2 + i\omega\gamma_0}. \qquad (46)$$

The average force on a water molecule is given by

$$F_{av3}v = T_m^{-1}\int_0^{T_m} dt\tilde{\lambda}\sum_j \delta(ja - vt)\dot{x}_j = T_m^{-1}(N\tilde{\lambda}^2/2mv^2)\sum_{\alpha}\int_{-\infty}^{\infty} d\omega \frac{\gamma}{(c^2/v^2 - 1)^2\omega^2 + \gamma_0^2}$$

$$= T_m^{-1}(N\tilde{\lambda}^2/2mv^2)/(c^2/v^2 - 1) \qquad (47)$$

giving, since c>>v,

$$F_{av3} \approx \frac{\tilde{\lambda}^2 N}{T_m mc^2 v} = \frac{\varepsilon'^2}{mc^2 a} \qquad (48)$$

for $T_m = Na/v$, $\tilde{\lambda} = \varepsilon'$.

The use of a delta function for the force exerted by a water molecule on the tube wall assumes that the range of the interaction does not extend beyond the carbon atoms that are opposite the water molecule under consideration. To approximately simulate a nonzero range interaction, let us replace the delta function by the following Lorentzian form, which reduces to the above delta function as $\Gamma \to 0$,

$$\frac{\tilde{\lambda}}{\pi} \frac{\Gamma}{(vt - ja)^2 + \Gamma^2} . \qquad (49)$$

$\Gamma$, which has units of length, is of the order of the range of the water-carbon interaction. Then, since

$$\frac{\tilde{\lambda}}{\pi} \int_{-\infty}^{\infty} dt' e^{i\omega t} \frac{\Gamma}{(vt' - ja)^2 + \Gamma^2} = \frac{\tilde{\lambda}}{\pi v} \int_{-\infty}^{\infty} dz e^{i\omega t} \frac{\Gamma}{(z - \omega ja/v)^2 + \Gamma^2} = v^{-1} \tilde{\lambda} e^{i(\omega/v)ja} e^{-\Gamma|\omega|/v}, \quad (50)$$

where $z = vt'$,

$$\dot{x}_j = \frac{\tilde{\lambda}}{mv} \sum_{k,\alpha} \int_{-\infty}^{\infty} d\omega \frac{i\omega e^{-i\omega t} e^{ikja} e^{-\Gamma|\omega|/v}}{(c^2/v^2 - 1)\omega^2 + i\gamma_0 \omega} . \qquad (51)$$

This gives

$$F_{av3} v = \frac{\tilde{\lambda}}{T_m} \sum_{k,\alpha} \int_{-T_m/2}^{T_m/2} dt \frac{1}{\pi} \frac{\Gamma}{(vt - ja)^2 + \Gamma^2} \dot{x}_j$$

$$\approx \frac{\tilde{\lambda}^2}{mv^2 T_m} \sum_{\alpha} \int_{-\infty}^{\infty} d\omega \frac{\gamma_0 e^{-2\Gamma|\omega|/v}}{(c^2/v^2 - 1)^2 \omega^2 + \gamma_0^2} = \frac{\tilde{\lambda}^2}{T_m mv^2 (c^2/v^2 - 1)} I' \approx \frac{\tilde{\lambda}^2}{T_m mc^2} I' \qquad (52)$$

for $\Gamma/v \ll T_m$, where

$$I' = \frac{1}{\pi} \int_{-\infty}^{\infty} du \frac{\gamma_0}{u^2 + \gamma_0^2} \exp\left(-\frac{2|u|\Gamma}{v(c^2/v^2 - 1)}\right) \qquad (53)$$

where $u = (c^2/v^2 - 1)\omega$. If $\Gamma \ll v(c^2/v^2 - 1)/(2\gamma_0)$, $I' \simeq 1$, and $F_{av3}$ is identical to the value given in Eq. (48). For $v \approx 1 m/s$, $c \approx 10^4 m/s$, $\gamma_0 \approx 10^{10} s^{-1}$, $I \simeq 1$, for $\Gamma \ll 10^{-2} m$, which is likely to be satisfied.

The product of the pressure and the area of the opening of the tube, which is equal to the rate at which momentum is imparted to the tube opening by water molecules in the reservoir, i.e., the net force on the chain, must be equal to $F_{av2}$ given in Eq. (17) for the water chain. The product of $F_{av3}$ and the number of water molecules in the tube for the case of non hydrogen bonded water molecules in the tube, which is probably a little smaller than $N_c$, is equal to the force acting on the water molecule at the end of the nanotube. Therefore, since

$$\frac{N_c F_{av3}}{F_{av2}} \sim \frac{a^{1/2} \gamma_0^{1/2} v^{1/2}}{c} , \qquad (54)$$

which is of the order of $10^{-4}$, Eq. (54) demonstrates that for $v$ of the order *1m/s* or less, the friction is noticeably smaller when the water molecules are not hydrogen bonded together.

## V. Conclusions

A simple model for the friction experienced by the one dimensional water chains that flow through subnanometer diameter carbon nanotubes is studied. The model is based on a lowest order perturbation theory treatment of the friction experienced by the water chains due to phonon and electron excitations in both the nanotube and the water chain, as a result of the motion of the water chain. Since the water chain is a one-dimensional material and the nanotube can behave as either a one-dimensional or two dimensional material, the friction due to phonon excitations has a contribution that is independent of the flow velocity and one which is a decreasing function of the flow velocity. Substituting the parameters used in molecular dynamics simulations in Eqs. (18) and (27), the friction due to phonon excitations for a flow velocity of the order of that reported in Ref. 9 was two orders of magnitude larger than the force acting on the water chain discussed in Ref. 9. This could be due either to the fact that the value that we used for the root mean square interaction $\lambda_0 = [<(\lambda_j - <\lambda_j>)^2>]^{1/2}$ was too large a fraction of $<\lambda_j>$ or the value that we used for $\varepsilon'$ was too large.

## Appendix A: An Estimate of the Rotational Free Energy of a Water Chain

What keeps the water chains in the middle of the nanotube is believed to be the free energy contribution due to the rotational entropy of the chain. This can be estimated from

$$f_r = -k_B T \ln Z, \qquad (A1)$$

where $f_r$ is the contribution to the free energy from the rotational motion of the chain and the partition function *Z* is given by

$$Z = \sum_{\ell=0}^{\infty} \exp[-\ell(\ell+1)\hbar^2 / (2Ik_B T)], \qquad (A2)$$

with the moment of inertia $I \approx MR^2$, where *R* is the mean radius of the chain. Substituting the parameters for the water chain, we get

$$Z = \sum_{\ell=1}^{\infty} \exp[-(1.93 \times 10^{-9})\ell(\ell+1)] \approx (2.28 \times 10^4) \int_0^{\infty} dx e^{-x^2} = 2.02 \times 10^4, \qquad (A3)$$

which gives $f_r = -0.248 eV$. Since $f_r$ is large in magnitude compared to $k_B T$, it is clear that the chain should stay in the middle of the nanotube, which will minimize its interaction with nanotube wall, which would suppress its rotation. This calculation is to an excellent approximation valid for the actual situation in which the nanotubes are long compared to the tube diameter and are curved over distances large compared to the tube diameter.

**Appendix B: Estimate of the electrical image potential force acting between the water chain and the carbon nanotube**

If the dipole moment of a water molecule points along the chain, it is given by $\vec{p} = p\hat{x}$. Its electrical potential is given by

$$V = \frac{1}{4\pi\varepsilon_0} \frac{px}{[x^2 + (z-z_0)^2]^{3/2}}, \qquad (B1)$$

where $z_0$ is its distance from the tube wall. Then the electrical image potential which must be added to this potential so that the resulting total potential vanishes at the wall (i.e., at z=0) is given by

$$V_i = -\frac{1}{4\pi\varepsilon_0} \frac{px}{[x^2 + (z+z_0)^2]}. \qquad (B2)$$

Then, the electrical image potential acting on the water molecule is given by

$$p \frac{\partial V_i}{\partial x}\bigg|_{x=0, z=z_0} = -\frac{1}{32\pi\varepsilon_0} \frac{p^2}{z_0^3} = -0.00125 eV \qquad (B3)$$

for p= the dipole moment of a water molecule and $z_0 = 3 \times 10^{-10} m$.

**Appendix C: Quantum Mechanical Treatment of Water Chain Flow through Nanotubes**

Let us look at a quantum mechanical treatment of this problem. The Steele potential has the form $V(x) = -\sum_j V_{0j} \cos[Q(x_j + vt)] \approx \sum_j \left[-V_{0j} \cos(Qvt) + \lambda_j x_j \sin(Qvt)\right]$, where $x_j$ is the displacement of a molecule due to this potential and $\lambda_j = QV_{0,j}$. A straightforward way to examine possible effects of quantum mechanics is to use Fermi's Golden rule with the above perturbation V(x). Then,

$$F_{av}v = \frac{\pi}{\hbar} \sum_{j,j'} <\lambda_j \lambda_{j'}> \sum_k \frac{\hbar}{2m\omega(k)N} e^{ik(j-j')a}$$
$$\times \left[\frac{\hbar\omega(k)e^{\beta\hbar\omega(k)}}{e^{\beta\hbar\omega(k)}-1}\left(\sum_{\pm}\delta(\hbar\omega(k)\pm\hbar Qv)\right) - \frac{\hbar\omega(k)}{e^{\beta\hbar\omega(k)}-1}\left(\sum_{\pm}\delta(\hbar\omega(k)\mp\hbar Qv)\right)\right] \qquad (C1)$$

Doing the average over the $\lambda$'s and making the approximation $\omega(k) \approx ck$, this expression becomes

$$F_{av}v = \frac{2\pi\lambda_0^2}{m} \frac{a}{2\pi} \int dk \left[\frac{e^{\beta\hbar c|k|}}{e^{\beta\hbar c|k|}-1}\delta(c|k|-Qv) - \frac{1}{e^{\beta\hbar c|k|}-1}\delta(c|k|-Qv)\right] = \frac{\lambda_0^2}{m}\frac{a}{c}. \qquad (C2)$$

Eq. (C2) is in agreement with Eqs. (13) and (17) for $\gamma_0 = 0$. The inclusion of phonon damping requires that we go beyond Fermi's Golden rule.

We can also use linear response theory. Let us first provide a short derivation of linear response theory: Consider a Hamiltonian $H = H_0 + H'$, where $H_0$ is the unperturbed Hamiltonian and $H'$ is a small perturbation. In the Schroedinger representation

$$i\hbar \frac{\partial}{\partial t} |n(t)> = H |n(t)>, \qquad (C3)$$

where $|n(t)>$ is the wave function of the system, which is the solution to Eq. (C3). Let $|n'(t)>$ represent this wave function in the interaction representation, where the transformation to the wave function in the interaction representation is given by

$$|n(t)> = e^{-iH_0 t/\hbar} |n'(t)>. \qquad (C4)$$

Substituting this expression in the Schroedinger equation, we get

$$i\hbar \frac{\partial}{\partial t} |n'(t)> = H'(t) |n'(t)>, \qquad (C5)$$

where

$$H'(t) = e^{iH_0 t/\hbar} H' e^{-iH_0 t/\hbar}, \qquad (C6)$$

whose solution is

$$|n'(t)> = \exp\left[-(i/\hbar)\int_{t_0}^{t} H'(t')dt'\right] |n'(t_0)>$$
$$\approx \left[1 - (i/\hbar)\int_{t_0}^{t} H'(t')dt'\right] |n'(t_0)> \qquad (C7)$$

to first order in $H'$. Consider the thermal average of A in the in the interaction representation,

$$<A>(t) = Z_0^{-1} \sum_{n'} e^{-\beta E_{n'}} <n'(t)|A(t)|n'(t)>, \qquad (C8)$$

where

$$A(t) = e^{iH_0 t/\hbar} A e^{-iH_0 t/\hbar}. \qquad (C9)$$

Then, from Eq. (C8)

$$<A>(t) \approx \sum_{n_0} e^{-\beta E_{n_0}} \left[<n'(t_0)|A(t)|n'(t_0)> - (i/\hbar)\int_{t_0}^{t} dt' <n'(t_0)|[H'(t'),A(t)]|n'(t_0)>\right], \qquad (C10)$$

where $H'$ is first turned on at $t = t_0$ and $n_0 = n'(t_0)$. Let $H'=Bf(t)$, where B is an operator and $f(t)$ is a scalar. Then

$$\delta <A>(t) = <A>(t) - <A>(t_0) = -(i/\hbar)Z_0^{-1}\sum_{n_0} e^{-\beta E_{n_0}} \int_{t_0}^{t} dt' f(t') <n_0 |[B(t'), A(t)]| n_0>. \quad (C10)$$

In our case, $A = \dot{x}_j$, $f(t) = \sin Qvt$ and $B = \sum_{j'} \lambda_{j'} x_{j'}$, giving

$$\delta <\dot{x}_j>(t) = <\dot{x}_j>(t) = -(i/\hbar)Z_0^{-1}\sum_{n_0} e^{-\beta E_{n_0}} \int_{t_0}^{t} dt' \sum_{j'} \lambda_{j'} \sin Qvt' <n_0 |[x_{j'}(t'), \dot{x}_j(t)]| n_0> \quad (C11)$$

since $<\dot{x}_j>(t_0) = 0$. The quantity

$$-(i/\hbar)\theta(t-t')Z_0^{-1}\sum_{n_0} e^{-\beta E_{n_0}} <n_0 |[x_{j'}(t'), \dot{x}_j(t)]| n_0> = \dot{G}_{j,j'}(t-t') \quad (C12)$$

is the derivative of the Green's function $G_{j,j'}(t-t')$, where $G_{j,j'}(t-t')$ satisfies

$$\frac{\partial^2}{\partial t^2} G_{j,j'}(t-t') - \sum_{j''} D_{j,j''} G_{j'',j'}(t-t') = \delta_{j,j'}\delta(t-t') \quad (C13)$$

where $D_{j,j'}$ is the dynamical matrix. The solution of Eq. (C13) is the expression for the Green's function given in Eq. (9).

The displacement and velocity operators are given by

$$x_j = a^{-1}\sum_{k,\alpha} \sqrt{\frac{\hbar}{m\omega_\alpha(k)}} (be^{ikja} + b^\dagger e^{-ikja}) \quad (C14a)$$

$$\dot{x}_j = a^{-1}\sum_{k,\alpha} \sqrt{\frac{\hbar\omega_\alpha(k)}{m}} (be^{ikja} - b^\dagger e^{-ikja}). \quad (C14b)$$

From the above expression for V, the force on the $j^{th}$ water molecule or carbon atom is equal to $-\partial V/\partial x_j = \lambda_j \sin Qvt$ giving for the power generated by the friction

$$F_{av}v = \lambda_j \sin Qvt \sum_{j'} <\dot{x}_{j'}> = \lambda_j \sin Qvt \int_{t_0}^{t} dt' \sum_{j'} \dot{G}_{j,j'}(t-t')\lambda_{j'} \sin Qvt' \quad (C15)$$

which when averaged over $\lambda_j \lambda_{j'}$ and t', gives us $F_{av1}v$. The phonon damping in the Green's function can be assumed to come from summing the diagrams for the phonon self-energy.

### Appendix D: Calculation of $\gamma'$ for a metallic nanotube

In Refs. [31,32], the expression for the friction coefficient $\lambda$ for a semiconducting nanotube is proportional to the integral

$$\sum_\alpha \sum_\pm \int_{-\infty}^{\infty} dk_i dk_f \frac{\hbar^2(k_f^2 - k_i^2)}{2m_\alpha} \delta\left[\frac{\hbar^2(k_f^2 - k_i^2)}{2m_\alpha} \pm \hbar v Q_x\right]\left[\exp\left(\frac{\mu - g_\alpha - \hbar^2 k_i^2}{k_B T}\right) - \exp\left(\frac{\mu - g_\alpha - \hbar^2 k_f^2}{k_B T}\right)\right]$$

(D1)

with $\mu \approx 0$, where $m_\alpha$ is the effective mass in the $\alpha^{th}$ band and $g_\alpha$ is the lowest energy of the $\alpha^{th}$ band. It is shown in Ref. [32] that this integral is equal to

$$\sum_\alpha I \frac{2m_\alpha v^2 Q_x^2}{k_B T} \exp\left(\frac{-g_\alpha}{k_B T}\right) \qquad (D2)$$

where $I$ is a dimensionless integral defined in Ref. [31]. For small radius nanotubes the $\alpha = 1$ dominates. The band gap $g_1$ is inversely proportional to the tube radius[41].

For a metallic nanotube, which is doped so that the Fermi level does not lie at zero wave vector, Eq. (D1) is replaced by

$$\sum_\alpha \sum_\pm \int_{-\infty}^{\infty} dk_i dk_f \frac{\hbar^2(k_f^2 - k_i^2)}{2m_0} \delta\left[\frac{\hbar^2(k_f^2 - k_i^2)}{2m_0} \pm \hbar v Q_x\right]\left[f(\hbar^2 k_i^2/2m_0) - f(\hbar^2 k_f^2/2m_0)\right] \quad (D3)$$

where

$$f(x) = \frac{1}{1 + \exp\left(\frac{x - \mu}{k_B T}\right)} \qquad (D4)$$

with $\mu \approx \varepsilon_F$, which to a good approximation can be treated as a step function. Eq. (D3) is easily shown to be equal to

$$\frac{m_0^2 Q_x^2 v^2}{\hbar^2 k_F^2}. \qquad (D5)$$

If the carbon nanotube is not doped, Eq. (D3) becomes

$$\sum_\alpha \sum_\pm \int_{-\infty}^{\infty} dk_i dk_f \hbar v_F (k_f - k_i) \delta\left[\hbar v_F (k_f - k_i) \pm \hbar v Q_x\right]\left[f(\hbar v_F k_i) - f(\hbar v_F k_f)\right], \qquad (D6)$$

where $v_F$ is the Fermi velocity, which is equal to

$$2\frac{v^2}{v_F^2} Q_x^2 \qquad (D7)$$

In order to get the friction coefficient for a metallic nanotube, we multiply the value of $\lambda$ in Appendix A of Ref. [32] by the ratio of Eq. (D5) to Eq. (D2), which is equal to

$$(3/8)(k_B T/\varepsilon_F)\exp(g_0/k_B T), \qquad (D8)$$

or the ratio of Eq. (D7) to Eq. (D2), which is equal to

$$\frac{k_B T}{2\varepsilon_F} \exp\left(\frac{g_0}{k_B T}\right). \qquad (D9)$$

**Acknowledgements:**

I thank Alex Noy for discussions useful that I had with him, and in particular for pointing out to me the likely role of rotational entropy of the water chain in keeping it in the center of the nanotube.